\documentclass{elsart5p}
\usepackage[dvips]{graphicx}
\usepackage{textcomp}
\usepackage{cite}
\usepackage{amsmath}
\usepackage{amsfonts}
\usepackage{booktabs}
\usepackage{amssymb}
\usepackage{ae}

\begin{document}

\begin{frontmatter}

\title{Development of cryogenic phonon detectors based~on~CaMoO$_{4}$~and~ZnWO$_{4}$~scintillating~crystals for~direct~dark~matter~search~experiments}

\author[MPI]{I. Bavykina\corauthref{cor1}}
\ead{bavykina@mppmu.mpg.de}
\author[MPI]{G. Angloher}
\author[MPI]{D. Hauff}
\author[MPI]{M. Kiefer}
\author[MPI]{F. Petricca}
\author[MPI]{F. Pr\"obst}

\address[MPI]{Max-Planck-Institut f\"ur Physik (Werner-Heisenberg-Institut), F\"ohringer Ring 6, 80805 Munich, Germany}
\corauth[cor1]{Corresponding author: tel: +49 89 323 54 316; fax: +49 89 323 54 526}

\received{6 June 2008}
\revised{23 September 2008}
\accepted{30 September 2008 }
\begin{abstract}
This work reports on the development of the first phonon detectors based on CaMoO$_{4}$ and ZnWO$_{4}$ scintillating crystals for the CRESST-II experiment. In particular, a novel technique for the production of the ZnWO$_{4}$ phonon detector with a separate thermometer carrier was investigated. The influence of the thermal and mechanical treatment on the scintillation light output of CaMoO$_{4}$~and~ZnWO$_{4}$ crystals at room temperature is discussed.
\end{abstract}
\end{frontmatter}

\section{Introduction}
The kinematics of galaxies and galaxy clusters suggests that the dynamic mass involved is far greater than the observable luminous mass \cite{Zwi33}. The detection of hypothetic new particles  constituting the missing mass of the Universe (the so-called Dark Matter) still remains one of the outstanding experimental efforts of present-day astrophysics and cosmology. In CRESST-II (second phase of the Cryogenic Rare Event Search with Superconducting Thermometers)  direct dark matter search experiments we are looking for WIMPs (Weakly Interacting Massive Particles) \cite{Jungman}. These particles are supposed to be gravitationally bound in a non-rotating roughly isothermal halo around the visible part of our galaxy with a density of about 0.3~GeV/cm$^{3}$ at the position of the Earth \cite{Caldwell}. We expect interaction with ordinary matter via elastic scattering on target nuclei. In case of a spin-independent  WIMP-nucleon scattering the interaction cross-section is proportional to A$^{2}$ (where A is atomic mass), thus favoring heavy nuclei in the target.

The present CRESST-II experimental setup at the Laboratori Nazionali del Gran Sasso (LNGS, Italy) involves a 66~channel SQUID readout system to enable operation of 33~detector modules which can house up to 10~kg of the target material \cite{CRESST, readout}.

\section{Conventional CRESST-II detector module}
The CRESST collaboration developed a reliable technique to produce sensitive cryogenic phonon detectors based on scintillating crystals \cite{Angloher}. When supplemented with a light detector  these provide an efficient discrimination of nuclear recoils from radioactive $\alpha$-, $\beta$- and $\gamma$- backgrounds while simultaneously measuring phonon and light signals caused by particle interactions in the scintillating crystal \cite{Meunier, Angloher1}. 

\begin{figure}[t]
\centering
\includegraphics[width=3.2in, clip=true]{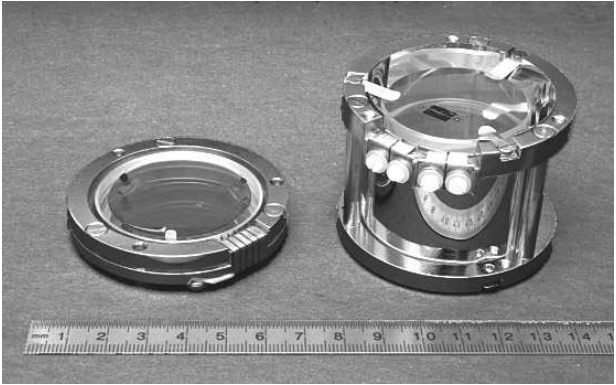}
\caption{An open CRESST-II detector module for coincident phonon and light measurements. On the left side: Silicon on sapphire light detector with W-SPT mounted in a copper holder. On the right side: Phonon detector based on a CaWO$_{4}$ scintillating crystal with W-SPT mounted in a copper holder. The crystal is held by six pairs of bronze clamps coated with an Al layer and with an Araldite~2011 scintillating layer\cite{Schafner}. The whole structure is surrounded by highly reflecting foil (VM2002).}
\label{fig:open_module}
\end{figure}

A conventional CRESST-II detector module consists of a cryogenic phonon detector based on a scintillating dielectric crystal and a nearby but separated light detector, optimized for the detection of scintillation light, see Figure~\ref{fig:open_module}. 

Each detector has a tungsten superconducting phase-transition thermometer (W-SPT) evaporated onto the surface. The thermometer layout is described in \cite{Angloher1}. We operate both detectors as cryogenic calorimeters detecting pulses on event by event basis while stabilizing the detectors within the superconducting-to-normal state of their transition curves. There a small temperature rise $\Delta$T of the W-SPT leads to a measurable rise $\Delta$R of its resistance, which can be read out with sensitive electronics. Since the shape and the steepness of the transition curve determines the dynamical range and the linearity of the detector response function, steep and narrow transition curves with transition temperatures as low as 20~mK are desirable.

To improve the collection of the scintillation light the whole structure is surrounded by a highly reflecting scintillating foil (VM2002)\footnote{VM2002 is a registered trademark of the 3M company. This foil has a sharp cutoff in reflectivity from $\sim$99~\% to almost 0~\%  for wavelengths lower than $\sim$390~nm at room temperature.}.

\subsection{Phonon detectors}
The copper housing is designed to hold scintillating crystals of cylindrical shape, 40~mm in diameter and 40~mm in height. All of our crystals (see Table \ref{table:properties}) have small $\sim$0.5~mm bevels. 
All surfaces are mechanically polished with silk to optical quality (with accuracy better then $\lambda$) except one flat surface which is roughened with 10~$\mu$m grain size. 
 \begin{table}[t]
\renewcommand{\arraystretch}{1.3}
\caption{Properties of CaWO$_{4}$, ZnWO$_{4}$ and CaMoO$_{4}$ scintillating crystals. The data are taken from \cite{Lecoq}; the light yield (LY) and scintillation decay time constant $\tau_{scint}$ of CaMoO$_{4}$ are taken from \cite{Korzhik}.}
\label{table:properties}
\begin{center}
\begin{tabular}{lccc}
\toprule
                                  &CaWO$_{4}$ & ZnWO$_{4}$  & CaMoO$_{4}$\\
\midrule
\midrule
Crystal structure                 &~scheelite~  & ~wolframite~  & ~scheelite~   \\
\midrule
Cleavage (plain)                    & ~weak (101)~&~marked (010)~&~weak (001)~   \\                             
\midrule
Index of refraction               &1.93       & 2.10        &1.98        \\
\midrule
Melting point ($^\circ$C)      & ~$\sim$1650~&~ $\sim$1200~  &~$\sim$1430~   \\
\midrule
Density (g/cm$^{3}$)             & 6.1        &7.8          &4.5            \\
\midrule
Emission max at 300~K             &420        &480          &520             \\
\midrule
Light Yield at 300~K           &6000         &21500          &   1720     \\
(ph/MeV) &&&\\
\midrule
$\tau_{scint}$ at 300~K ($\mu$s)            &0.6        &22         &20             \\
\midrule
Debye frequency (THz)              &$\sim$4.7 & $\sim$3.8   &$\sim$8.2        \\

\bottomrule
\end{tabular}
\end{center}
\end{table}

The absence of a noticeable degradation of the light yield for events near the crystal surface is essential for our crystals to be operated as phonon detectors at low temperatures~\cite{Meunier}. Therefore, chemically stable and non-hygroscopic crystals are favored as scintillating targets. Since the crystals are thoroughly held by six pairs of metal clamps in the holder, it is important that they are mechanically stable.

In the CRESST-II experiment we use calcium tungstate crystals as target material \cite{Angloher1}. Among different scintillators CaWO$_{4}$ was selected because of its relatively high light yield at low temperature\footnote{The light yield of the scintillating crystal at low temperatures determines the discrimination threshold between nuclear and electron recoils.}. Due to the need of CRESST-II to further improve the capability of rejecting background caused by known particles scintillating crystals with a light yield higher than that of CaWO$_{4}$ are constantly under investigation. Additionally, for the interpretation of the positive WIMP signature, it is beneficial to have targets with a variety of nuclei. 

\subsection{Light detectors}
The scintillating light signal is measured using a separate light detector made either from a sapphire wafer (40~mm in diameter and $\sim$0.4~mm thick) with an epitaxially grown silicon absorber layer, or made from a silicon wafer with a similar shape (40~mm in diameter and $\sim$0.4~mm thick) of square (30$\times$30$\times$0.45)~mm$^{3}$ shape \cite{Petricca}.

\section{Phonon detectors based on CaMoO$_{4}$ crystals} 
\begin{figure}[t]
\centering
\includegraphics[width=3.2in, clip=true]{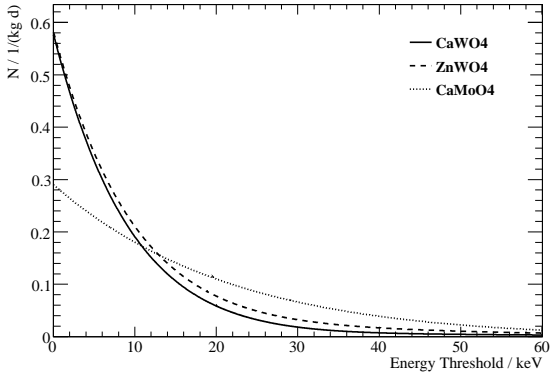}
\caption{The integrated recoil spectra expected for 100~GeV/c$^{2}$ WIMPs scattered elastically on CaWO$_{4}$ (solid line), ZnWO$_{4}$ (dashed line) and CaMoO$_{4}$ (dotted line) targets. Coherent WIMP-nucleon scattering cross-section is 1~pb. The used dark matter profile is described in Section~1.}
\label{fig:intdedn}
\end{figure}

The comparison between the WIMP-nucleon interaction rate on different target materials would give us unique possibility to verify a positive WIMP signal and to learn about WIMP properties. For example, on a CaMoO$_{4}$ target WIMPs are expected to scatter mainly on molybdenum nuclei\footnote{This is due to the factor A$^{2}$ in the spin-independent scattering cross-section.} whereas on a CaWO$_{4}$ or a ZnWO$_{4}$ target they would mainly scatter on tungsten nuclei. Figure~\ref{fig:intdedn} shows the integrated recoil spectra plotted as a function of the discrimination threshold energy expected for 100~GeV/c$^{2}$ WIMPs scattering on various scintillating targets with a spin-independent WIMP-nucleon scattering cross-section of 1~pb~\cite{Raffi1}. The count rate on a CaMoO$_{4}$ target above a discrimination threshold of  $\sim$14~keV is higher than that on a CaWO$_{4}$ target. In a CaWO$_{4}$ target most of the WIMP induced recoils are expected to appear below 40~keV, while in a CaMoO$_{4}$ target this value is shifted towards a higher energy of $\sim$60~keV.    

The first characterization of the light yield of the CaMoO$_{4}$ crystals at a temperature of 8~K proofed the potential of these scintillating crystals as target material. It was measured that the light yield increases progressively with decreasing temperature \cite{Mikhailik}. We are currently measuring the light yield of a CaMoO$_{4}$ sample at~mK temperatures in an assembly similar to the conventional CRESST-II detector module (see Section~2).

Last but not least, CaMoO$_{4}$ is an excellent candidate to search for neutrinoless double-$\beta$ decay (0$\nu$2$\beta$) of $^{100}$Mo~\cite{Annenkov, Pirro}. The integration of phonon detectors based on CaMoO$_{4}$ crystals into the CRESST-II experiment would give us the interesting possibility to perform a multi-experiment with only one target material and to obtain a competitive upper limit on the half-life of 0$\nu$2$\beta$ decay of $^{100}$Mo.  

\subsection{Relative light yield of CaMoO$_{4}$ crystals at room temperature}
Two cylindrically chopped (40~mm in diameter and 40~mm in height) crystals considered in this work were produced by the Czochralsky technique at Bogoroditsk Plant of Technochemical Products (BTCP, Bogoroditsk, Tula region, Russia) from purified raw materials. The relative light yield (RLY) at room temperature was measured using a green extended PMT Photonis (XP3461B)\footnote{The quantum efficiency of a Photonis XP3461B is maximal around 420~nm.} with a base modified to integrate scintillation light over 400~$\mu$s. The crystals were mounted in a plastic holder of 50~mm in diameter. The wall of the holder was covered with a reflecting foil (VM2002). Crystals were measured without optical coupling to the PMT and were irradiated by 662~keV $\gamma$'s from a $^{137}$Cs source. For each crystal the measured light yield with applied spectral correction~\cite{corrections} was normalized to the light yield of our standard CaWO$_{4}$ crystal ($\O$=40~mm, height=40~mm). The results of the RLY measurements of CaMoO$_{4}$ crystals ($\O$=40~mm, height=40~mm) at room temperature are summarized in Table~\ref{table:CaMoO}. The energy resolution was estimated as the FWHM divided by the position of the 662~keV line.     
\begin{table}[t]
\renewcommand{\arraystretch}{1.3}
\caption{Relative light yield of CaMoO$_{4}$ crystals measured at room temperature with a PMT. The light yield of each CaMoO$_{4}$ crystal with applied spectral correction was normalized to the light yield of our standard CaWO$_{4}$ crystal (see text for details).}
\label{table:CaMoO}
\begin{center}
\begin{tabular}{lcc}
\toprule

                       &~RLY~($\%$) ~& ~Resolution~($\%$)\\   
&&                                                   at 662~keV\\                                                   
\midrule
\midrule
CaMoO$_{4}$-1 as arrived        & 17        & 33.7   \\
\midrule
CaMoO$_{4}$-1 annealed          &15.5       & 37.6  \\                             
\midrule 
CaMoO$_{4}$-1 with W-SPT,       & 20.4      & 26        \\
annealed, roughened&&\\
\midrule
CaMoO$_{4}$-2 with W-SPT,       & 22        &23.7        \\
roughened, no annealing &&\\
\bottomrule
\end{tabular}
\end{center}
\end{table}

\begin{figure}[t]
\centering
\includegraphics[width=3.2in, clip=true]{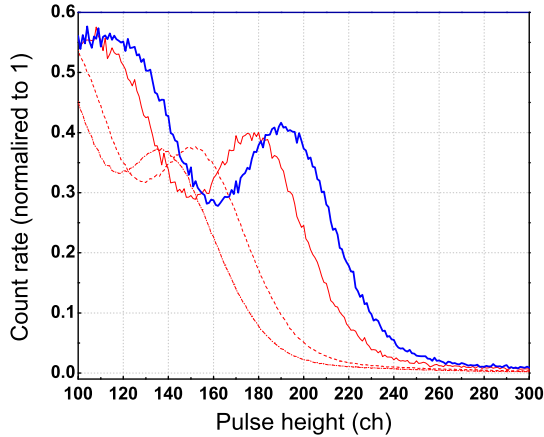}
\caption{$^{137}$Cs spectra measured with the CaMoO$_{4}$-1 crystal before annealing (dashed line), after annealing at 800~$^\circ$C~ for 48 hours in oxygen flow (dash dotted line), and after W-SPT deposition with subsequent surface roughening (thin solid line), together with $^{137}$Cs spectrum measured with a different crystal CaMoO$_{4}$-2 after W-SPT deposition with subsequent surface roughening (thick solid line). Measurements were performed at room temperature with a PMT (see text for details).}
\label{fig:CaMoO}
\end{figure}

Since after-growth annealing at high temperatures can have a significant influence on the scintillation properties of the crystals we performed a series of tests to investigate this issue. Annealing at 800~$^\circ$C~for 48~hours in oxygen flow\footnote{It was not known at this time that the producer annealed the crystal CaMoO$_{4}$-1 at $\sim$1300~$^\circ$C~in oxygen flow.} degraded the light yield and the energy resolution of the crystal CaMoO$_{4}$-1 to 15.5~$\%$ and to 37.6~$\%$, respectively (see Fig.~\ref{fig:CaMoO}). Since smaller (20$\times$10$\times$5)~mm$^{3}$ CaMoO$_{4}$ samples were not annealed in oxygen flow by the producer they have shown an increase of the RLY after annealing performed at 800~$^\circ$C~for 48~hours in oxygen flow.

Roughening of one of the flat crystal surfaces with boron carbide powder which followed W-SPT deposition (see Section~3.2) improved the light yield and the energy resolution to 20.4~$\%$ and 26~$\%$, respectively. This increase was expected since more scintillating photons were able to escape through the roughened surface facing the PMT.      

\subsection{W-SPT on CaMoO$_{4}$ crystals}
For the W-SPT deposition a crystal was mounted in a tantalum holder (which can be kept at the desired deposition temperature via radiative heating) into the tungsten deposition system. Before deposition an area below a thermometer on the CaMoO$_{4}$ crystal surface was roughened on atomic level with an ion-gun. This procedure improves the adhesion of the film to the crystal surface. Before the deposition the crystal was slowly heated up to 450~$^\circ$C~ in vacuum and then kept at this temperature during the whole deposition process. To prevent inter-diffusion between the tungsten film and the CaMoO$_{4}$ crystal surface we deposited 2~k$\r{A}$ layer of SiO$_{2}$. Then a 2~k$\r{A}$ layer of tungsten was evaporated. There was no crystal quality degradation observed after W-SPT deposition. 

\begin{figure}[t]
\centering
\includegraphics[width=2.0in, clip=true]{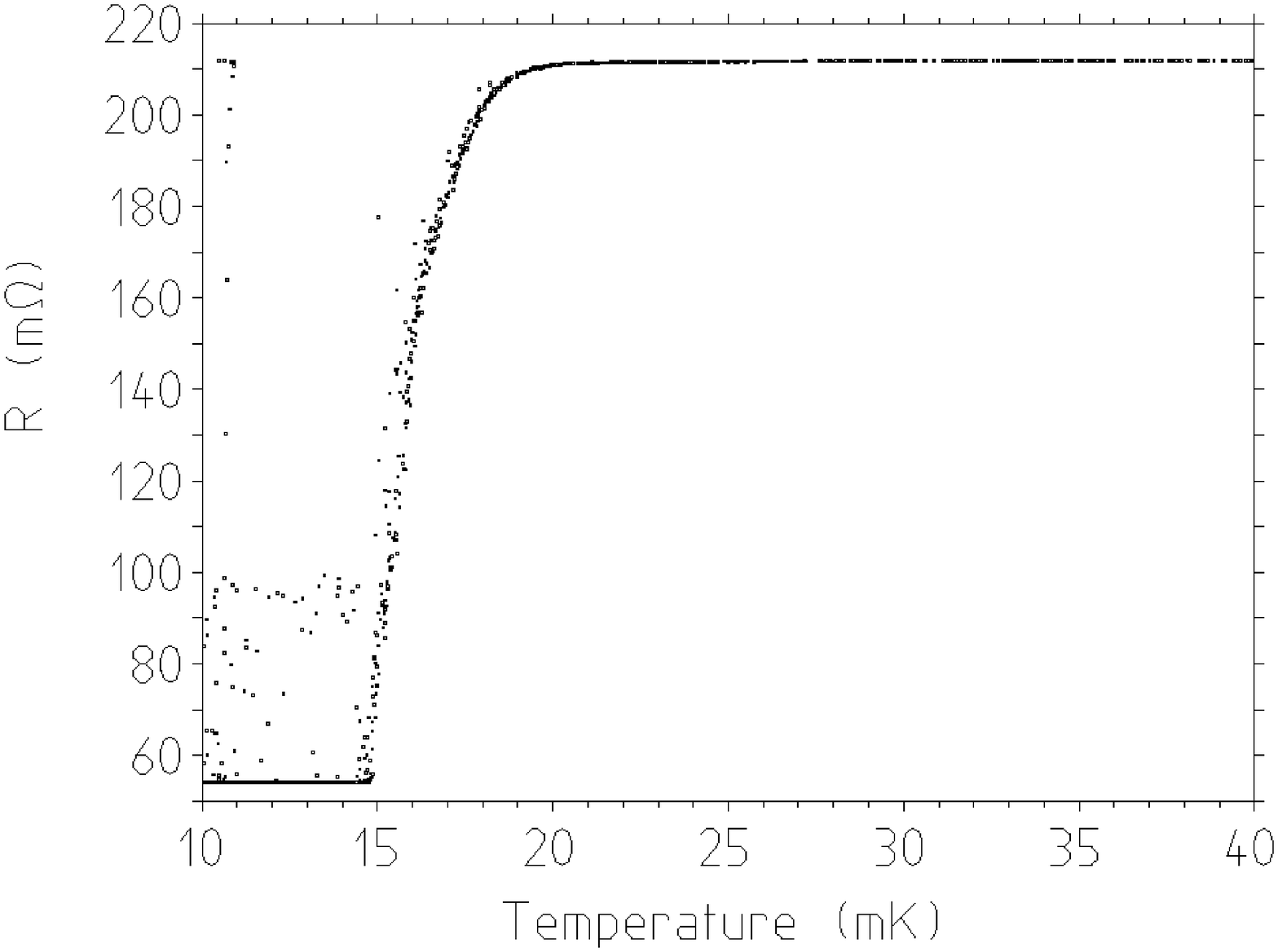}
\includegraphics[width=2.0in, clip=true]{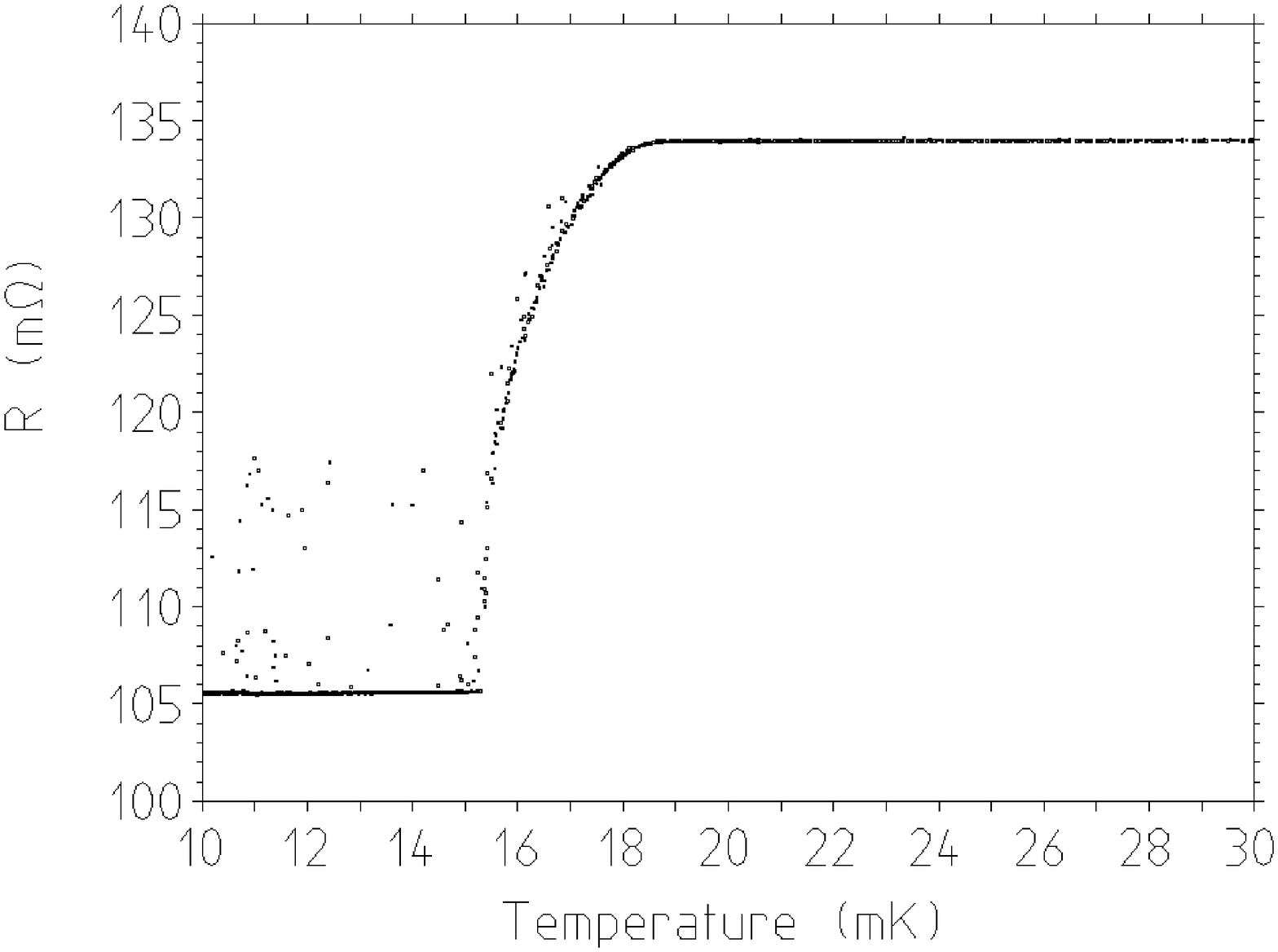}
\caption{The transition curves of W-SPT on the CaMoO$_{4}$-1 crystal (top) and of W-SPT on the different CaMoO$_{4}$-2 crystal (bottom) recorded with a 1~$\mu$A bias current.   }
\label{fig:CaMoOtrans}
\end{figure}

The resistance of the thermometer was measured by passing a constant bias current through the readout circuit in which the thermometer was in parallel with a small shunt resistor and an input coil of a dc-SQUID. While cooling to the base temperature of a dilution fridge a decrease in the thermometer resistance was measured as a current decrease through the SQUID input coil (for details see \cite{Angloher1}). The resistances between $\sim$130~m$\Omega$ and $\sim$200~m$\Omega$ of the W-SPT on the CaMoO$_{4}$ crystals were measured as described above while applying~1~$\mu$A bias current. Steep and narrow transition curves with transition temperatures of $\sim$18~mK and width of a few~mK were achieved, see Fig.~\ref{fig:CaMoOtrans}. The characteristics of the obtained transition curves fulfill the quality requirements of the W-SPT for the CRESST-II phonon detectors. 

The thermal response \cite{Franz} of the W-SPT on the CaMoO$_{4}$-1 crystal which was stabilized in the normal-to-superconducting state has shown promising results. However, due to the high count rate on ground level ($\sim$50~Hz) no calibration spectrum was recorded. 

\section{Phonon detectors based on ZnWO$_{4}$ crystals} 
In search of scintillators with high light yield at~mK temperatures we also analyzed ZnWO$_{4}$ crystals in detail. The results of the research have shown that the relative light yield of ZnWO$_{4}$ at low temperatures is comparable with that of CaWO$_{4}$ \cite{Bavykina, Kraus}. Additionally, the count rate from ZnWO$_{4}$ crystals due to internal radioactive impurities is expected to be lower than that from CaWO$_{4}$ crystals \cite{Probst}. Thus, ZnWO$_{4}$ appears to be an attractive target for the CRESST-II dark matter search and the first prototype phonon detector based on this material was developed.    
 
\subsection{Relative light yield of ZnWO$_{4}$ crystals at room temperature}
All ZnWO$_{4}$ crystals considered in this work were produced by the unique low temperature gradient Czochralsky technique \cite{Novosib} at Nikolaev Institute of Inorganic Chemistry (NIIC, Novosibirsk, Russia). The results of the spectroscopic analysis of one of the ZnWO$_{4}$ crystal measured with a high purity germanium detector in the Garching underground laboratory (Techniche Universit\"at M\"unchen) are summarized in Table~\ref{table:ZnWOspectroscopy}. Since the density of the crystal is rather high, the germanium detector efficiencies for small $\gamma$ energies are rather low, thus leading to higher upper limits for the activities.
\begin{table}[t]
\renewcommand{\arraystretch}{1.3}
\caption{Upper limits on the activities of the ZnWO$_{4}$-3 crystal ($\O$=40~mm, height=16~mm, mass=163~g) for the different isotopes from the $^{232}$Th-, the $^{238}$U- and the $^{235}$U-decay chain, as well as for $^{40}$K. When several $\gamma$ lines were used for the calculation of the activity of one isotope, the quoted value is the weighted average of the single activities. The limits given for the whole decay chain are again the weighted averages of the activities of the single isotopes \cite{Hofmann}.}
\label{table:ZnWOspectroscopy}
\begin{center}
\begin{tabular}{lc}
\toprule

                      Isotope~& ~Activity (90$\%$C.~L.)\\ 

\midrule
\midrule
K-40                  &   $\textless$321.3$^{+29.8}_{-26.2}$~mBq/kg\\
\midrule
$^{228}$Ac             &   $\textless$124.6$^{+8.0}_{-7.1}$~mBq/kg \\
$^{212}$Pb             &   $\textless$315.7$^{+50.7}_{-47.2}$~mBq/kg\\
$^{212}$Bi             &   $\textless$266.5$^{+18.3}_{-16.0}$~mBq/kg\\
\midrule
Th-232 chain    &   $\textless$151.1$^{+7.3}_{-6.5}$~mBq/kg\\
\midrule
$^{234}$Pa           &   $\textless$562.3$^{+52.0}_{-45.7}$~mBq/kg\\
$^{226}$Ra           &   $\textless$29.9$^{+2.8}_{-2.5}$~Bq/kg\\
$^{214}$Pb           &   $\textless$115.4$^{+8.7}_{-7.7}$~mBq/kg\\
$^{214}$Bi           &   $\textless$159.4$^{+10.5}_{-9.3}$~mBq/kg\\
\midrule
U-238 chain    &   $\textless$140.5$^{+6.6}_{-5.9}$~mBq/kg\\
\midrule
$^{235}$U           &   $\textless$3.61$^{+0.33}_{-0.29}$~Bq/kg\\
$^{227}$Th           &   $\textless$1.20$^{+0.25}_{-0.24}$~Bq/kg\\
$^{211}$Pb           &   $\textless$161.5$^{+16.7}_{-15.9}$~mBq/kg\\
\midrule
U-235 chain    &   $\textless$175.7$^{+16.7}_{-15.9}$~mBq/kg\\

\bottomrule
\end{tabular}
\end{center}
\end{table}

\begin{figure}[t]
\centering
\includegraphics[width=3.2in, clip=true]{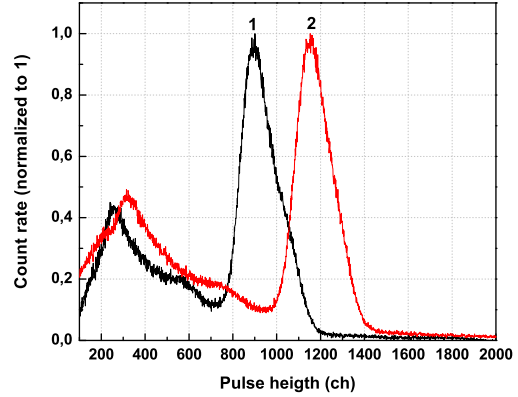}
\caption{$^{137}$Cs spectra measured with the roughened ZnWO$_{4}$-4 crystal before annealing (1) and after annealing at 800~$^\circ$C~ for 48 hours in oxygen flow (2). Measurements were performed at room temperature with a PMT (see text for details).}
\label{fig:ZnWoO4}
\end{figure}

The relative light yield of the crystals was measured in the experimental setup described in Section 3.1 and followed the same analysis procedure as in case of CaMoO$_{4}$ crystals. The results of the measurements are summarized in Table~\ref{table:ZnWO}.   
\begin{table}[t]
\renewcommand{\arraystretch}{1.3}
\caption{Relative light yield of ZnWO$_{4}$ crystals measured at room temperature with a PMT. The light yield of each ZnWO$_{4}$ crystal with applied spectral correction was normalized to the light yield of our standard CaWO$_{4}$ crystal (see text for details).}
\label{table:ZnWO}
\begin{center}
\begin{tabular}{lcc}
\toprule

                               &~RLY~($\%$) ~& ~Resolution~($\%$)\\ 
        &&                                                   at 662 keV\\ 
\midrule
\midrule
\multicolumn{3}{l}{ZnWO$_{4}$-1 ($\O$=40~mm, height=10~mm, mass=92.5~g)} \\
\midrule
as arrived                                      & 95.4        & 19.3   \\
\\
annealing 1 at 800$^\circ$C~for 24h          &101       & 16.7  \\                             
\\
roughened after annealing 1& 117     & 10.8        \\
\\
annealing 2 at 900$^\circ$C~for 48h, &107&10.8\\
roughened &&\\
\midrule
\midrule
\multicolumn{3}{l}{ZnWO$_{4}$-3 ($\O$=40~mm, height=16~mm, mass=163~g, roughened)} \\
\midrule
as arrived                                      & 118       & 12.5  \\
\midrule
\midrule
\multicolumn{3}{l}{ZnWO$_{4}$-4 ($\O$=40~mm, height=40~mm, mass=404.4~g, roughened)} \\
\midrule
as arrived                           & 83.6        & 15.7   \\
annealed at 800$^\circ$C~for 48h         &111.1       & 16  \\                            
 
\bottomrule
\end{tabular}
\end{center}
\end{table}
In general, annealing at 800~$^\circ$C~for 48~hours in oxygen flow substantially improved RLY of the ZnWO$_{4}$ crystals. This increase can be explained by the deficit of oxygen content in the delivered crystals which was compensated by the annealing in oxygen flow\footnote{The emission of the luminescence light is associated with the radiative transition between tungsten and oxygen within the (WO$_{6}$)$^{6-}$ molecular  complex.}. Figure~\ref{fig:ZnWoO4} shows $^{137}$Cs spectra measured with ZnWO$_{4}$-4 crystal before and after annealing by PMT at room temperature (see Section~3.1). From this figure it is evident that the decomposition into two Gaussians gives a much better fit to a measurement than the fit to a single Gaussian as would be expected. This effect can be partially explained with geometrical trapping of scintillation light due to the total internal reflections at the crystal surfaces \cite{Ninkovich}. Probably because of the increased transparency of the crystal (which has an impact on the light propagation through the crystal) a double structure of the $^{137}$Cs photopeak almost disappeared after applying the annealing procedure. Further investigations of the influence of annealing on the scintillation properties of ZnWO$_{4}$ crystals are in progress.   

Roughening of the flat surface of the crystal with boron carbide powder further improved the RLY and energy resolution, thus additionally reducing trapping of the scintillation light within the crystal.   

\begin{figure}[t]
\centering
\includegraphics[width=3.2in, clip=true]{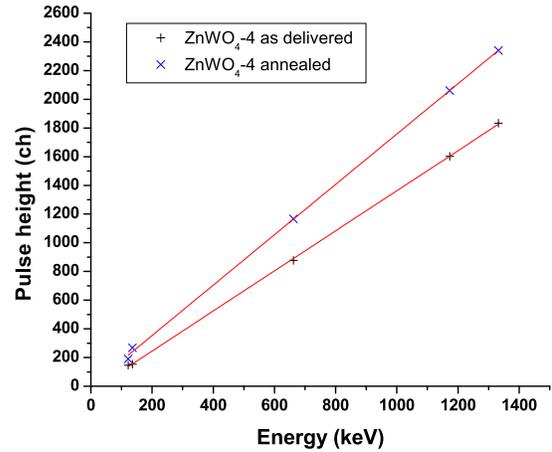}
\caption{Linearity of the scintillation response of the ZnWO$_{4}$-4 crystal before and after annealing at 800~$^\circ$C~for 48~ hours in oxygen flow. Measurements were taken via irradiation of the crystal with $\gamma$'s from  $^{57}$Co (122~keV and 136~keV), $^{137}$Cs (662~keV) and $^{60}$Co (1.17~MeV and 1.33~MeV) calibration sources. Measurements were performed at room temperature with a PMT (see text for details).} 
\label{fig:linearity}
\end{figure}

A good linearity of the scintillation response between 122~keV and 1.33~MeV was measured with the ZnWO$_{4}$-4 crystal before and after annealing, see Fig.~\ref{fig:linearity}. 

\subsection{W-SPT on ZnWO$_{4}$ crystals}
The research performed in \cite{Bavykina1} has shown that very high deposition temperatures of $\sim$740~$^\circ$C~are needed in order to obtain transition temperatures of $\sim$22~mK for W-SPT on ZnWO$_{4}$ crystals. Heating up a crystal to such a high temperature in vacuum degrades the scintillation output of a crystal dramatically. Therefore, we decided to grow a W-SPT on a separate thermometer carrier and to apply a novel gluing technique \cite{Kiefer, Jean} to produce a phonon detector based on a ZnWO$_{4}$ crystal (see Section 4.3).

A~(20$\times$10$\times$1)~mm$^{3}$ ZnWO$_{4}$ wafer of known orientation was mounted into the tungsten deposition system and slowly heated up to 600~$^\circ$C~ in vacuum, and kept at this temperature during the whole deposition process. To prevent inter-diffusion between the tungsten film and the ZnWO$_{4}$ crystal surface we first deposited a 2~k$\r{A}$ layer of SiO$_{2}$. Then, a 2~k$\r{A}$ layer of tungsten was evaporated. The $\sim$120~m$\Omega$ resistance of the W-SPT on the ZnWO$_{4}$ wafer was measured as described in Section~3.2 while applying~50~nA bias current. Transition temperatures as low as $\sim$19~mK were achieved using the  deposition procedure described above. The characteristics of the obtained transition curve fulfill the quality requirements of the W-SPT for the CRESST-II phonon detectors.

\subsection{ZnWO$_{4}$ phonon detector with a separate thermometer carrier}
Since ZnWO$_{4}$ crystals tend to easily cleave along the (010) plane, we first investigated the influence of the glue layer on the surface properties of the crystals. Two different glues, namely Araldite~2011 and Epo-Tek 301-2\footnote{Araldite~2011 is a registered trademark of the Huntsman Corporation. Epo-Tek~301-2 is a registered trademark of the Epoxy Technology.}, and two different thermometer carrier crystals (CaWO$_{4}$ and ZnWO$_{4}$) were tested. In each test, the carrier crystal was glued with a certain glue type on a ZnWO$_{4}$ crystal ($\O$=40~mm, height=10~mm). This composite structure was then mounted in a copper holder and cooled down to~mK temperatures. The results of these measurements are summarized in Table~\ref{table:glue}.      
\begin{table}[t]
\renewcommand{\arraystretch}{1.3}
\caption{Gluing of different thermometer carriers on a ZnWO$_{4}$ crystal. After cooling to mK temperatures, surface quality of some crystals remained unchanged (--), whereas degradation was observed for others (+), see text for details.}
\label{table:glue}
\begin{center}
\begin{tabular}{lcc}
\toprule

                                  &CaWO$_{4}$ wafer ~& ~ZnWO$_{4}$ wafer~\\ 
& ~(20$\times$10$\times$1)~mm$^{3}$~&~(20$\times$10$\times$1)~mm$^{3}$~\\ 
\midrule
\midrule
Epo-Tek 301-2                                     & ~--~        & ~--~   \\
Araldite 2011   &~+~  & ~+~ \\                             

\bottomrule
\end{tabular}
\end{center}
\end{table}
Only in case of gluing with Epo-Tek~301-2 surface quality of crystals remained unchanged, whereas degradation was observed in case of gluing with Araldite~2011. 

\begin{figure}[t]
\centering
\includegraphics[width=2.0in, clip=true]{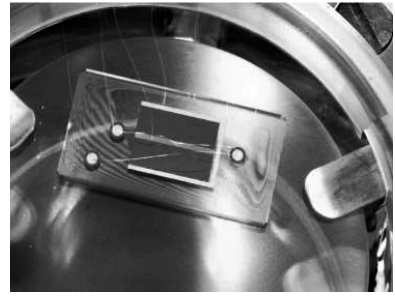}
\caption{ZnWO$_{4}$ phonon detector with the separate ZnWO$_{4}$ thermometer carrier mounted in a copper holder. The crystal is held by six pairs of bronze clamps coated with an Al layer and with an Araldite~2011 scintillating layer. The whole structure is surrounded by a highly reflecting foil VM2002. The bond wires for the electrical contacts are visible.}
\label{fig:glued}
\end{figure}

The (20$\times$10$\times$1)~mm$^{3}$ ZnWO$_{4}$ wafer of known orientation (plane~(010) is perpendicular to the (20$\times$10)~mm$^{2}$ surface of the wafer) with the W-SPT on it (see Section~4.2) was glued with Epo-Tek 301-2 on the ZnWO$_{4}$-4 target crystal of know orientation (plane~(010) is perpendicular to the flat surface of the crystal). Figure~\ref{fig:glued} shows the produced ZnWO$_{4}$ phonon detector with the glued thermometer carrier mounted in a copper holder.   

\section{Conclusions}  
In the framework of the CRESST-II experiment we developed reliable techniques for the production of cryogenic phonon detectors based on CaMoO$_{4}$ and ZnWO$_{4}$ scintillating crystals. The light output of CaMoO$_{4}$ and ZnWO$_{4}$ crystals at room temperature have been studied. Furthermore, the influence of the mechanical (polishing, roughening) and thermal treatment (annealing, W-SPT deposition) on the light output of these scintillating crystals has been investigated. 

The developed prototype ZnWO$_{4}$ phonon detector with a separate glued on thermometer was installed into the main CRESST-II experimental setup at Gran Sasso underground laboratory for the ongoing dark matter Run~31 and has already shown promising results.  The scintillation response of the CaMoO$_{4}$ crystals at low temperatures is the subject of the ongoing investigation.

\end{document}